\documentclass[prl,aps,superscriptaddress,showpacs,twocolumn]{revtex4}
\usepackage{graphicx}

 \def\Slash#1{
  \begin{picture}(5,6)(0,0)
  \put(-.7,-1.2){\line(5,6)6}
  \end{picture}
  \kern-.8em#1}
 \def\slash#1{
  \begin{picture}(5,6)(0,0)
  \put(-1.5,-1.7){\line(5,6)5}
  \end{picture}
  \kern-.8em#1}

\def\gg5{\gamma_5}
\def\hg5{\hat{\gamma}_5}
\def\g4{\gamma_4}

\def\Qlatmr1{Q_{lat}^{(m=r=1)}}

\def\be{\begin{eqnarray}}
\def\ee{\end{eqnarray}}

\def\gmu{\gamma_{\mu}}

\def\g5{\gamma_5}

\usepackage{amssymb}

\begin{document}
 

\title{Theoretical Foundation for the Index Theorem on the Lattice 
with Staggered Fermions}

\author{David H.~\surname{Adams}}
\email{dhadams@ntu.edu.sg}

\affiliation{Division of Mathematical Sciences, Nanyang Technological
University, Singapore 637371}
\thanks{Current and permanent address}

\affiliation{NCTS, National Taiwan University, Taipei, Taiwan}

\date{Dec.~15, 2009}

\begin{abstract}

A way to identify the would-be zero-modes of staggered lattice
fermions away from the continuum limit is presented. Our approach
also identifies the chiralities of these modes, and their index
is seen to be determined by gauge field topology in accordance with the
Index Theorem. 
The key idea is to consider the spectral flow of a certain 
hermitian version of the staggered Dirac operator.
The staggered fermion index thus obtained can be used as a new way to
assign the topological charge of lattice gauge fields.
In a numerical study in U(1) backgrounds in 2 dimensions it is found to 
perform as well as the Wilson index while being computationally more 
efficient. It can also be expressed as the index of an overlap Dirac 
operator with a new staggered fermion kernel.

\end{abstract}


\maketitle

The computational efficiency and other attractive features of the staggered
lattice fermion formulation made possible the first high-precision lattice 
simulations of Quantum Chromodynamics (QCD) \cite{Davies(PRL)}, and it
continues to be widely used.
However, this formulation has long been perceived as disadvantaged 
 compared to Wilson fermions -- the other traditional lattice fermion 
formulation -- regarding topological aspects of QCD. 
This concerns in particular the Index Theorem \cite{AS(index)} relating gauge 
field topological charge to fermionic zero-modes, which is needed
to explain the large mass of the $\eta$' meson \cite{Witten-V}. 
A field-theoretic approach to the staggered fermion index was 
developed and used by Smit and Vink \cite{SV,SV(eta-top)};
however, it does not give an integer value for the index from the beginning
and requires a renormalization that depends on the full ensemble of lattice
gauge fields. 
In contrast, for Wilson fermions, an integer-valued index is obtained without
any need for renormalization. This is because Wilson fermions have identifiable
would-be zero-modes with definite chirality, provided the lattice gauge field
is not too rough, from which the index can be defined.  
The Wilson fermion index \cite{SV,Itoh,overlap(H)}, calculated from 
its equivalent description as the index of the overlap Dirac operator 
\cite{Neu(overlap)}, is currently widely used for issues such as 
calculation of the topological susceptibility 
\cite{DelDebbio} and investigations of topological structure in the QCD vacuum 
-- see, e.g., \cite{Fritz(new)}. 
But for staggered fermions it has not been known how to identify the would-be 
zero modes and their chiralities in non-smooth gauge field backgrounds; they 
were only seen to emerge in simulations with improved actions that are 
sufficiently close to the continuum \cite{Davies-Wong-Durr}.

In this paper we show how the would-be chiral zero-modes of staggered 
fermions can be identified away from the continuum limit when the background
gauge field is not too rough, thus determining an integer-valued index.
A theoretical foundation for the Index Theorem for staggered fermions is 
established, placing them on the same footing as Wilson fermions in this 
regard.

Our new approach in the staggered case parallels the {\em spectral flow} 
approach to the Index Theorem for continuum and Wilson lattice fermions 
\cite{Itoh,overlap(H)}, which we begin by briefly reviewing in the following.
Spacetime is taken to be a Euclidean box with periodic boundary conditions
(the setting in which Lattice QCD simulations are performed) 
with even dimension $d$. Besides the physically relevant case $d=4$ we will 
also consider $d=2$ for illustrative purposes. 
From the hermitian Dirac gamma matrices $\{\gmu\}_{\mu=1,\dots,d}$ the 
chirality matrix $\gamma_5=-(i)^{d/2}\gamma_1\cdots\gamma_d$ is defined.
It has the properties $\gamma_5^{\dagger}=\gamma_5$, $\gamma_5^2={\bf 1}$, 
$\{\gamma_5\,,\gamma_{\mu}\}\!=\!0$ and $\{\gamma_5\,,D\}\!=\!0$ 
where $D=\gmu(\partial_{\mu}+A_{\mu})$ is the continuum massless Dirac 
operator on spinor fields coupled to a gauge field $A$. 
Consequently, the vectorspace of zero-modes of $D$ (i.e. solutions to 
$D\psi=0$) decomposes into $\pm$ chirality subspaces on which 
$\gamma_5=\pm{\bf 1}$. The {\em index} of $D$ is the difference between
the numbers $n_{\pm}$ of independent $\pm$ chirality zero-modes, and is
fixed by topology: 
Gauge fields with smooth field tensor $F_{\mu\nu}(x)$ have an integer
topological charge $Q$, 
and the Index Theorem in this setting states 
\be
n_+-n_-=(-1)^{d/2}Q\,.
\label{2}
\ee
The {\em spectral flow} perspective on the index arises by considering the 
eigenvalues $\{\lambda(m)\}$ of the hermitian operator
\be
H(m)=\gamma_5(D-m)
\label{3}
\ee
as a function of the parameter $m$.
Note that a zero-mode $\psi$ of $D$ with $\pm$ chirality
is also an eigenmode of $H(m)$ with eigenvalue 
$\lambda(m)=\mp m$, crossing the origin with slope $\mp1$ at $m=0$.
Furthermore, from the property   
\be
H(m)^2=D^{\dagger}D+m^2
\label{4}
\ee
we see that these are the {\em only} eigenvalues of $H(m)$ that cross the 
origin at any value of $m$. 
It follows that the spectral flow of $H(m)$, defined as the net 
number of eigenvalues $\lambda(m)$ of $H(m)$ that cross the origin, counted 
with sign $\pm$ depending on the slope of the crossing, comes entirely from 
eigenvalue crossings at $m=0$ and equals $n_--n_+$, i.e. minus the index.
 
In the lattice setting with Wilson fermions,
the would-be zero-modes can be identified as the low-lying {\em real} 
eigenvalues of the Wilson-Dirac operator $D_W$ \cite{SV}.
The spectral flow perspective \cite{Itoh,overlap(H)} is
based on the hermitian lattice analogue of (\ref{3}):
\be
H_W(m)=\gamma_5(D_W-m)
\label{6}
\ee
Regarding the spectral flows $H_W(m)\psi(m)=\lambda(m)\psi(m)$, note that
$\lambda(m_0)=0\;\Leftrightarrow\;D_W\psi(m_0)=m_0\psi(m_0)$.
Thus eigenvalue crossings of $H_W(m)$ are in one-to-one correspondence
with {\em real} eigenvalues of $D_W$. 
Furthermore, after normalizing eigenmodes such that $\psi^{\dagger}\psi=1$ 
one easily finds $\lambda'(m)=-\psi(m)^{\dagger}\gamma_5\psi(m)$ \cite{Itoh}.
Thus the sign of the slope of $\lambda(m)$ at a low-lying crossing value 
$m_0$ is minus the chirality of the corresponding would-be 
zero-mode $\psi(m_0)$ for $D_W$. It follows that the index of $D_W$  
is minus the spectral flow of $H_W(m)$ coming from the 
eigenvalues which cross the origin at low-lying values of $m$. 
Numerical results illustrating this can be found, e.g., in \cite{Itoh}.
An illustration in the $d\!=\!2$ case is given in Fig.~4 below.

Turning now to staggered lattice fermions,
where the lattice field $\chi(x)$ is scalar (rather than spinor) and
describes $2^{d/2}$ degenerate continuum fermion species (called quark tastes),
the massless staggered Dirac operator is 
\be
D_{st}=\eta_{\mu}\nabla_{\mu}
\label{9}
\ee
where $\nabla_{\mu}$ is the usual lattice finite difference operator coupled
to the lattice gauge field and 
$\eta_{\mu}\chi(x)=(-1)^{n_1+\dots+n_{\mu-1}}\chi(x)$  where 
$x=a(n_1,\dots,n_d)$ runs over the lattice sites. 
$D_{st}$ is anti-hermitian and therefore has purely imaginary spectrum.
Consequently the identification of would-be zero-modes in the Wilson case
does not carry over to the staggered case since it relied crucially on the
role of non-zero real eigenvalues. For the same reason it is clear that 
an index of $D_{st}$ cannot be obtained from spectral flow of a staggered 
version of $H_W(m)$. In fact the staggered analogue of (\ref{6}), 
$\Gamma_5(D_{st}-m)$, is not even hermitian. However, there is an alternative 
spectral flow approach which we now discuss, and which turns out to be 
perfectly suited to the staggered case.

Return momentarily to the continuum setting and note that
$H(m)$ in (\ref{3}) is not the only hermitian operator that can be used
for the spectral flow perspective on the index. 
We could just as well use $H(m)=iD-m\gamma_5$.
Its spectral flow is equal to minus the index of $D$ just as before, since
the previous argument, including the property (\ref{4}), holds verbatim for
this operator. But now the analogue in the staggered case,
\be
H_{st}(m)=iD_{st}-m\Gamma_5
\label{10}
\ee
is also hermitian, so we can consider its spectral flow as well.
Here $\Gamma_5$ is the analogue of $\gamma_5$ in the 
staggered formulation; it is hermitian and corresponds up to $O(a)$ 
discretization errors to $\gamma_5\otimes{\bf 1}$ in the spin-flavor 
interpretation \cite{Golterman-Smit}.
Note that $H_{st}(0)=iD_{st}$, so the eigenvalues of $D_{st}$ are 
$\{-i\lambda(0)\}$ where $\{\lambda(m)\}$ are the eigenvalue flows
of $H_{st}(m)$. This fact allows us to identify the would-be 
zero-modes of $D_{st}\,$: as shown below, they are the eigenmodes with 
eigenvalues $-i\lambda=-i\lambda(0)$ for which the associated flow $\lambda(m)$ 
crosses zero at a low-lying value of $m$. Furthermore, the sign of the slope 
of the crossing is minus the chirality of the would-be zero-mode, and hence 
the index is minus the spectral flow of $H_{st}(m)$ coming from the crossings
at low-lying values of $m$. 

To see this, consider first the situation in a smooth continuum-like gauge 
field background: The eigenvalues $-i\lambda$ of the would-be zero-modes 
$\chi$ of $D_{st}$ separate out from the rest of the spectrum; they are almost 
zero and have approximately definite chirality 
$\chi^{\dagger}\Gamma_5\chi\approx\pm1$.
For the corresponding eigenvalue flows $\lambda(m)$ of $H_{st}(m)$
we have $\lambda'(m)=-\chi(m)^{\dagger}\Gamma_5\chi(m)$ just as in the Wilson
case. Since $\chi=\chi(0)$ it follows that $\lambda'(0)\approx\mp1$, and since 
$\lambda=\lambda(0)$ is almost zero it then follows that $\lambda(m)$ crosses 
the origin at a very small (in magnitude) value $m_0$ of $m$. 
Whether $m_0$ is positive 
or negative depends on the chirality and sign of $\lambda(0)$, but in either 
case the sign of the slope of the crossing is the same as that of 
$\lambda'(0)$, i.e. minus the chirality. Under a roughening of the 
gauge field the location of a crossing will move, but it cannot
disappear until it meets another crossing with opposite slope. Hence the
would-be zero-modes remain identifiable with unchanged chiralities and index.

To illustrate the identification of the staggered would-be zero-modes and 
their index we present results of a numerical study in U(1) backgrounds
in 2 dimensions. Following \cite{SV} we 
start from specific smooth lattice gauge fields with topological
charge $Q$ and roughen them by multiplying the link variables by 
random phase factors: $U_{\mu}(x)\to e^{ir_{\mu}(x)\delta}U_{\mu}(x)$ with each
$r_{\mu}(x)$ randomly chosen in $[-\pi,\pi]$; the parameter
$\delta$ controls the roughness. 
Our numerical computations were all on the 12 x 12 lattice (as in \cite{SV}) 
and done using ARPACK \cite{Arpack}. The code was checked by reproducing the 
results in Table 1 of \cite{SV}.

Figure 1 shows low-magnitude eigenvalues of $H_{st}(m)$ versus $m$
(horizontal axis)
in a $Q\!=\!1$ background of moderate roughness $\delta=0.33$. 
The low-magnitude eigenvalues of $iD_{st}$ are the eigenvalues 
at $m\!=\!0$ in the figure. There is no clear separation in their magnitudes.
Nor is there a clear separation in the magnitudes of the chiralities 
as measured by $\chi^{\dagger}\Gamma_5\chi$: numerical calculation 
of this quantity for the 3 pairs of eigenmodes with lowest magnitude 
eigenvalues gives $-0.28,\;+0.17,\;+0.13$.
Nevertheless, the would-be zero-modes can now be identified among these 
low-magnitude modes: as discussed above, they are precisely the eigenmodes 
whose eigenvalues $\lambda=\lambda(0)$ belong to eigenvalue flows 
$\lambda(m)$ which cross zero at a low-lying value of $m$. 
From the figure we see that there are two of these, both with positive slope,
corresponding to negative chirality. Thus the index is $-2$. This is in 
accordance with the Index Theorem, since in the staggered fermion case
(\ref{2}) becomes
\be
\mbox{index}(D_{st})=2^{d/2}(-1)^{d/2}Q 
\label{11}
\ee
As a second illustration, Figure 2 shows the eigenvalue flow in a $Q\!=\!-2$
background of the same roughness level $\delta=0.33$.
From the eigenvalues at $m\!=\!0$ we see that 
$iD_{st}\!=\!H_{st}(0)$ has one eigenvalue pair of much smaller magnitude
than the others. Numerical calculation of $\chi^{\dagger}\Gamma_5\chi$
for these gives $+0.34$ compared to $+0.21,\;+0.13$
for the eigenmode pairs with the next 2 lowest-magnitude eigenvalues.
Hence we would naively expect the first pair to be the would-be zero-modes, 
giving index $+2$.
However, from the eigenvalue flows we see that there is in fact
one more pair of would-be zero-modes of $D_{st}$, also with positive chirality, 
giving index $+4$ in accordance with the Index Theorem (\ref{11}).    
\begin{figure}
\includegraphics[width=2.8in,clip]{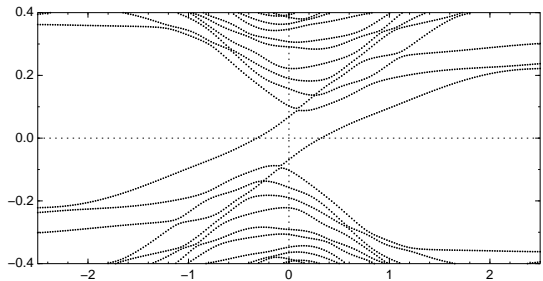}
\caption{\label{sf1eps} Staggered spectral flow 
in a $Q\!=\!1$ background.}
\end{figure} 
\begin{figure}
\includegraphics[width=2.8in,clip]{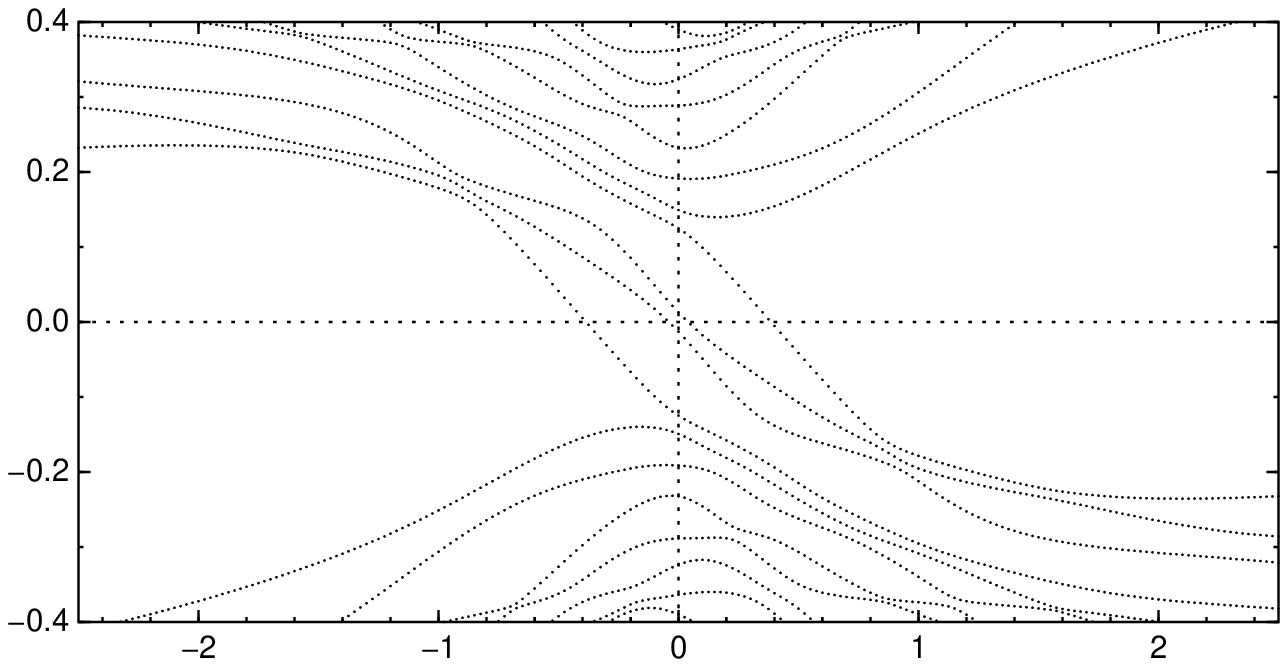}
\caption{\label{sf2eps} Staggered spectral flow 
in a $Q\!=\!-2$ background.}
\end{figure} 
\begin{figure}
\includegraphics[width=2.8in,clip]{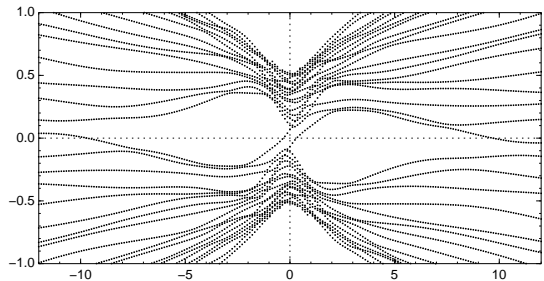}
\caption{\label{sf3eps} Staggered spectral flow over a larger $m$ range 
in same background as Fig.~1.}
\end{figure} 
\begin{figure}
\includegraphics[width=2.8in,clip]{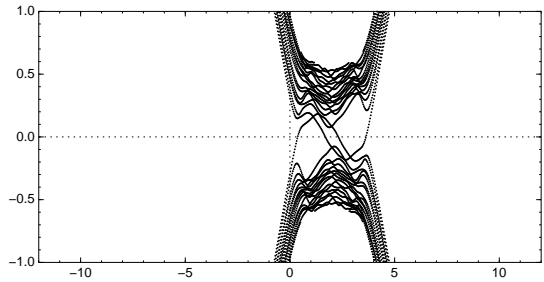}
\caption{\label{sf4eps} Wilson spectral flow in same background as 
Fig's 1,3.}
\end{figure} 

A crucial property that would-be zero-modes of a lattice fermion formulation
should have is {\em robustness}: they should not disappear under small 
deformations of the gauge field. This is assured in the present case if
there is a clear separation between low- and high-lying crossing regions.
Figure 3 shows the eigenvalue flow of $H_{st}(m)$ in the same background as 
Fig.~1 but over a larger $m$ range. 
We see that the eigenvalue crossings only occur in a localized
region around $m=0$ and in high-lying regions, $|m|\gtrsim9$ in this case.
The large separation between low-lying and high-lying crossings illustrates
the robustness of the would-be zero-modes and index for staggered 
fermions. For comparison,
Figure 4 shows the eigenvalue flow of the hermitian Wilson 
operator $H_W(m)$ in the same gauge field background.
It has one low-lying positive-slope crossing in accordance
with the Index Theorem, and the high-lying crossings are localized 
around $m=2$ and $m=4$ as expected on theoretical grounds 
\cite{Itoh,L-Neu-DA(bound)}.

As in the Wilson case \cite{L-Neu-DA(bound)}, separation between 
low- and high-lying 
crossing regions can be proved analytically when the plaquette variables of 
the lattice gauge field satisfy the approximate smoothness condition 
$||1-U_{\mu\nu}(x)||<\epsilon$ 
for sufficiently small $\epsilon\,$\cite{values}: 
We derive in \cite{DA(stoverlap)} a bound of the form 
\be
H_{st}(m)^2\;\ge\;
\left\{
\begin{array}{ll}
m^2-K(m,d)\epsilon & \quad \mbox{for $|m|\le1$}\\
1-K(m,d)\epsilon & \quad \mbox{for $|m|\ge1$}\\
\end{array} \right.
\label{bound}
\ee
(which is found to be saturated in the free field case where $\epsilon=0$).
The precise form of $K(m,d)\ge0$ is not important here, only the fact that it
depends continuously on $m$, which ensures that $K_0$ defined in the following 
is finite. 
For any $b_1,b_2$ with $0<b_1<1<b_2$
set $K_0=\mbox{max}\{K(m,d)\;|\;|m|\le b_2\}$ and 
$\epsilon_0=\frac{b_1^2}{K_0}$, then the bound (\ref{bound}) implies
$H_{st}(m)^2>0$ for $b_1\le|m|\le b_2$ when $\epsilon<\epsilon_0$ in the 
plaquette condition. This shows that the separation between low-lying 
($|m|<b_1$) and high-lying ($|m|>b_2$) eigenvalue crossing regions can be 
made arbitrarily large by taking $\epsilon>0$ to be sufficiently small.

The performance of the staggered index compared to the Wilson index was
investigated in the numerical study with $Q\!=\!1$. For each randomly generated
$\{r_{\mu}(x)\}$ we examined the would-be zero-modes and index as the 
roughness parameter $\delta$ is increased. In both the staggered and Wilson
cases they were found to remain identifiable up to roughness levels 
$0.33\le\delta\le0.41$ (which easily exceeds the limit $\delta\approx0.25$
at which the field-theoretic approach to the staggered fermion index
was found to break down in \cite{SV}). 
Furthermore, in all the backgrounds considered, the
breakdown value of $\delta$ for staggered was found to be the same
as for Wilson up to differences $\Delta\delta=\pm0.02$ (which favored 
staggered as often as Wilson). This suggests that
the staggered and Wilson indexes are accessing the same topological content 
of the lattice gauge field. However, the computational cost was roughly twice 
as much in the Wilson case. This is as expected since $d\!=\!2$ Wilson 
fermions have two spinor components.

Finally we show that, analogously to the Wilson case, the staggered fermion 
index can be obtained as the index of the exact zero-modes of an overlap Dirac 
operator $D_{ov}$. This is of practical as well as theoretical interest since
the Wilson fermion index is usually calculated in practice as 
$\mbox{index}(D_{ov})$. 
The role of $\gamma_5$ in the overlap construction 
\cite{Neu(overlap)} is not to be played by $\Gamma_5$ here since
it violates the required property $\gamma_5^2=1$ by an $O(a)$ term.
Instead we use
\be 
\Gamma_{55}\chi(x)=(-1)^{n_1+\dots+n_d}\chi(x),
\label{28}
\ee
which corresponds to $\gamma_5\otimes\gamma_5$ in the spin-flavor 
interpretation \cite{Golterman-Smit}. We define the overlap Dirac 
operator with staggered fermion kernel by
\be
D_{ov}=\frac{1}{a}\Big(1+\Gamma_{55}\frac{H_{st}(m_0)}{\sqrt{H_{st}(m_0)^2}}
\Big)\,.
\label{29}
\ee
As in the Wilson case \cite{Neu(overlap)} $D_{ov}$ has exact zero-modes, with 
definite chirality with respect to $\Gamma_{55}$, and satisfying an index 
formula
\be
\mbox{index}(D_{ov})
=-\frac{1}{2}\mbox{Tr}\Big(\frac{H_{st}(m_0)}{\sqrt{H_{st}(m_0)^2}}\Big)\,. 
\label{30}
\ee
which also follows from the general index formula in \cite{Has(index)}
after noting that (\ref{29}) satisfies the Ginsparg-Wilson relation
\cite{Has(index)} with $\gamma_5\to\Gamma_{55}$. 
As in the Wilson case we take $m_0$ to be in the region in between the 
positive low-lying and high-lying eigenvalue crossings of $H_{st}(m)$,
then (\ref{30}) gives 
\be
\mbox{index}(D_{ov})={\textstyle \frac{1}{2}}\,\mbox{index}(D_{st})
\label{31}
\ee
To see this, note that $H_{st}(m)\Gamma_{55}=-\Gamma_{55}H_{st}(-m)$, which
implies a symmetry in the eigenvalue flow: $H_{st}(m)$ and $-H_{st}(-m)$
have the same spectrum (as seen in Figures 1,2,3). It follows that $H_{st}(0)$
has symmetric spectrum; therefore (\ref{30}) is minus the spectral flow
from $m=0$ to $m=m_0$, which in turn is minus half the spectral flow 
from $-m_0$ to $m_0$, i.e. half of $\mbox{index}(D_{st})$.

The staggered overlap operator introduced here is of independent interest
as a new fermion formulation for Lattice QCD, and will be studied
in a separate paper \cite{DA(stoverlap)}. It has the remarkable feature
of reducing the number of staggered fermion tastes by half (as reflected
in the factor $1/2$ in (\ref{31})). The physical fields turn out to correspond
to the two continuum tastes with positive flavor-chirality under
${\bf 1}\otimes\gamma_5$, so that 
$\gamma_5\otimes\gamma_5$ chirality is the same as the physical 
$\gamma_5\otimes{\bf 1}$ for them. 
A new staggered version of domain wall fermions is also
obtained \cite{DA(stoverlap)}.

In summary,
staggered lattice fermions do maintain the important Index Theorem connection 
between gauge field topology and  fermionic zero-modes, but in a way that was
not realized previously. In the present $d=2$ study it was seen to perform
as well as the Wilson index, but with greater numerical efficiency.
Future work should investigate the performance of the index for improved
staggered fermions versus Wilson index in backgrounds generated in 
current Lattice QCD simulations in 4 dimensions. 


\begin{thebibliography}{XXX}
\bibitem{values} As a practical illustration, 
$\max_x|1-U_{\mu\nu}(x)|$ for the backgrounds of Fig.'s
1 and 2 is 1.949 and 1.950, respectively, which is very 
close to the maximum possible value 2.
\end{thebibliography}
\end{document}